\documentclass{pasj00}

\def\submitted{submitted}
\def\inpress{in press}

\def\arxiv#1{ (arXiv astro-ph/#1)}

\DeclareAbbreviation\AAHam{Astron. Abh. Hamburg. Sternw.}
\DeclareAbbreviation\AARv{Astron. Astrophys. Rev.}
\DeclareAbbreviation\AAS{American Astron. Soc. Meeting Abstracts}
\DeclareAbbreviation\AcA{Acta Astron.}
\DeclareAbbreviation\actaa{Acta Astron.}
\DeclareAbbreviation\Afz{Astrofizika}
\DeclareAbbreviation\AGAb{Astronomische Gesellschaft Abstract Ser.}
\DeclareAbbreviation\an{Astron. Nachr.}
\DeclareAbbreviation\AnAp{Annales d'Astrophysique}
\DeclareAbbreviation\AnTok{Tokyo Astron. Obs. Annals, Sec. Ser.}
\DeclareAbbreviation\Ap{Astrophysics}
\DeclareAbbreviation\ARep{Astron. Rep.}
\DeclareAbbreviation\AstBu{Astrophys. Bull.}
\DeclareAbbreviation\ATel{Astron. Telegram}
\DeclareAbbreviation\ATsir{Astron. Tsirk.}
\DeclareAbbreviation\AcApS{Acta Astrophys. Sinica}
\DeclareAbbreviation\AstL{Astron. Lett.}
\DeclareAbbreviation\BaltA{Baltic Astron.}
\DeclareAbbreviation\BANS{Bull. of the Astron. Institutes of the Netherlands Suppl. Ser.}
\DeclareAbbreviation\BASI{Bull. Astron. Soc. India}
\DeclareAbbreviation\BeSN{Be Newslett.}
\DeclareAbbreviation\BHarO{Harvard Coll. Obs. Bull.}
\DeclareAbbreviation\CBET{Cent. Bur. Electron. Telegrams}
\DeclareAbbreviation\CEAB{Central European Astrophys. Bull.}
\DeclareAbbreviation\ChJAA{Chinese J. of Astron. and Astrophys.}
\DeclareAbbreviation\caa{Chinese J. of Astron. and Astrophys.}
\DeclareAbbreviation\CoAsi{Asiago Contr.}
\DeclareAbbreviation\CoSka{Contributions of the Astronomical Observatory Skalnat\'e Pleso}
\DeclareAbbreviation\GCN{GRB Coord. Netw. Circ.}
\DeclareAbbreviation\ErgAN{Erg. Astron. Nachr.}
\DeclareAbbreviation\ibvs{IBVS}
\DeclareAbbreviation\IEEEP{IEEE Proc.}
\DeclareAbbreviation\JAD{J. Astron. Data}
\DeclareAbbreviation\JApA{J. of Astrophys. and Astron.}
\DeclareAbbreviation\JAVSO{J. American Assoc. Variable Star Obs.}
\DeclareAbbreviation\JBAA{J. Br. Astron. Assoc.}
\DeclareAbbreviation\JPhCS{J. of Physics Conference Series}
\DeclareAbbreviation\JPSJ{J. Phys. Soc. Japan}
\DeclareAbbreviation\JSARA{J. of the Southeastern Assoc. for Research in Astron.}
\DeclareAbbreviation\LowOB{Lowell Obs. Bull.}
\DeclareAbbreviation\MitAG{Mitteil. der Astronom. Gesell. Hamburg}
\DeclareAbbreviation\MitVS{Mitteil. Ver\"{a}nderl. Sterne}
\DeclareAbbreviation\MmSAI{Mem. Soc. Astron. Ital.}
\DeclareAbbreviation\memsai{Mem. Soc. Astron. Ital.}
\DeclareAbbreviation\Msngr{Messenger}
\DeclareAbbreviation\NewA{New Astron.}
\DeclareAbbreviation\na{New Astron.}
\DeclareAbbreviation\NewAR{New Astron. Rev.}
\DeclareAbbreviation\nar{New Astron. Rev.}
\DeclareAbbreviation\NInfo{Nauchnye Informatsii}
\DeclareAbbreviation\NPhS{Nature Physical Science}
\DeclareAbbreviation\OAP{Odessa Astron. Publ.}
\DeclareAbbreviation\Obs{Observatory}
\DeclareAbbreviation\OEJV{Open Eur. J. on Variable Stars}
\DeclareAbbreviation\PASA{Publ. Astron. Soc. Australia}
\DeclareAbbreviation\PASAu{Publ. Astron. Soc. Australia}
\DeclareAbbreviation\PAZh{Pis'ma AZh}
\DeclareAbbreviation\PJAB{Proc. Japan Acad. Ser. B}
\DeclareAbbreviation\POBeo{Publ. de l'Observatoire Astronomique de Beograd}
\DeclareAbbreviation\PCCP{Phys. Chem. Chem. Phys.}
\DeclareAbbreviation\PhR{Phys. Rep.}
\DeclareAbbreviation\PVSS{Publ. Variable Stars Sect. R. Astron. Soc. New Zealand}
\DeclareAbbreviation\PZ{Perem. Zvezdy}
\DeclareAbbreviation\PZP{Perem. Zvezdy, Prilozh.}
\DeclareAbbreviation\QJRAS{QJRAS}
\DeclareAbbreviation\RA{Ricerche Astronomiche}
\DeclareAbbreviation\RMxAA{Rev. Mexicana Astron. Astrof.}
\DeclareAbbreviation\RvMA{Reviews of Modern Astron.}
\DeclareAbbreviation\RvMP{Reviews of Modern Phys.}
\DeclareAbbreviation\SASS{Society for Astronom. Sciences Ann. Symp.}
\DeclareAbbreviation\Sci{Science}
\DeclareAbbreviation\SPIE{SPIE Proc.}
\DeclareAbbreviation\SvA{Soviet Astronomy}
\DeclareAbbreviation\SvAL{Soviet Astronomy Letters}
\DeclareAbbreviation\VeSon{Ver\"{o}ff. Sternw. Sonneberg}
\DeclareAbbreviation\VSOLJBul{VSOLJ Variable Star Bull.}
\DeclareAbbreviation\yCat{VizieR Online Data Catalog}
\DeclareAbbreviation\ZA{Z. Astrophys.}

\def\PublisherCambridge{Cambridge: Cambridge University Press}

\begin{document}
\SetRunningHead{T. Kato and Y. Osaki}{Negative Superhumps in KIC 7524178}

\Received{201X/XX/XX}%{yyyy/mm/dd}
\Accepted{201X/XX/XX}%{yyyy/mm/dd}

\title{KIC 7524178 -- an SU UMa-Type Dwarf Nova Showing Predominantly 
Negative Superhumps throughout Supercycle}

\author{Taichi \textsc{Kato}}
\affil{Department of Astronomy, Kyoto University,
       Sakyo-ku, Kyoto 606-8502}
\email{tkato@kusastro.kyoto-u.ac.jp}

\and

\author{Yoji \textsc{Osaki}}
\affil{Department of Astronomy, School of Science, University of Tokyo,
Hongo, Tokyo 113-0033}
\email{osaki@ruby.ocn.ne.jp}

%%% end:list of authors


\maketitle

\begin{abstract}
We analyzed the Kepler long cadence data of KIC 7524178
(=KIS J192254.92$+$430905.4), and found
that it is an SU UMa-type dwarf nova with frequent normal outbursts.
The signal of the negative superhump was always the dominant one
even during the superoutburst, in contrast to our common
knowledge about superhumps in dwarf novae.
The signal of the positive superhump was only transiently seen
during the superoutburst, and it quickly decayed after the
superoutburst.  The frequency variation of the negative superhump
was similar to the two previously studied dwarf novae in the
Kepler field, V1504 Cyg and V344 Lyr.
This is the first object in which the negative 
superhumps dominate throughout the supercycle.  Nevertheless,
the superoutburst was faithfully accompanied by the positive superhump,
indicating that the tidal eccentric instability is essential for 
triggering a superoutburst.  All the pieces of evidence strengthen 
the thermal-tidal instability as the origin of the superoutburst
and supercycle, making this object the third such example in 
the Kepler field.  This object had unusually small ($\sim$1.0 mag)
outburst amplitude and we discussed that the object has
a high mass-transfer rate close to the thermal stability limit
of the accretion disk.  The periods of the negative and positive
superhumps, and that of the candidate orbital period were 0.07288~d
(average, variable in the range 0.0723--0.0731~d), 0.0785~d
(average, variable in the range 0.0772--0.0788~d) and 
0.074606(1)~d, respectively.
\end{abstract}

\section{Introduction}

   Cataclysmic variables (CVs) are close binary systems
consisting of a white dwarf and a red-dwarf secondary 
transferring matter via the Roche-lobe overflow
[for a review, see \citet{war95book}].
SU UMa-type dwarf novae are a class of CVs which show
superoutbursts in addition to normal dwarf-nova outbursts.
The most distinguishing feature of superoutbursts is
the presence of superhumps (positive superhumps), whose periods
are a few percent longer than the orbital period.
The origin of the superhump is generally considered as
the consequence of the growth of the tidal eccentric
instability when the disk radius reaches the 3:1 resonance
\citep{whi88tidal}, and the increased tidal dissipation 
results in the long, bright superoutburst (\cite{osa89suuma};
\cite{osa96review}) [thermal-tidal instability (TTI) model].
Although the TTI model has been widely accepted, it has been 
challenged by an irradiation-induced mass-transfer model
(e.g. \cite{sma91suumamodel}; \cite{sma04EMT};
\cite{sma09SH}) and a pure thermal instability model
(e.g. \cite{can10v344lyr}).  Quite recently, 
\citet{osa13v1504cygKepler} demonstrated that the variation
of the frequency of the negative superhumps, which have shorter
periods than the orbital period and are considered as
a result of a tilted disk (e.g. \cite{har95v503cyg};
\cite{pat97v603aql}; \cite{woo07negSH}), exactly reproduces
the disk radius variation predicted by the TTI model.

   KIC 7524178 (also referred to as KIS J192254.92$+$430905.4,
hereafter KIS J1922) is a CV identified as an object
with an H$\alpha$ emission \citep{sca13CVKeplerfield}.
KIS J1922 showed relatively strong Balmer and He\textsc{I}
emission lines on a blue continuum \citep{sca13CVKeplerfield}.
These emission lines were sharp, suggesting a low orbital
inclination.  We analyzed the long cadence (LC) public Kepler data 
(quarters 15 and 16) of this object and found that it is
an SU UMa-type dwarf nova with short outburst intervals.
What is most surprising was that negative superhumps
were predominantly present even during the superoutburst.
In this Letter, we report our analysis of the object and
examine whether such an unusual phenomenon
can be understood within the framework of the TTI model.

\section{Data Analysis and Results}

   The data analysis was performed practically in the same way
as in \citet{osa13v1504cygKepler} and \citet{osa13v344lyrv1504cyg}:
we used two-dimensional Fourier analysis with and without
the Han window function, and least absolute shrinkage and selection
operator (Lasso; \cite{Lasso}; \cite{kat12perlasso};
\cite{kat13j1924}) for detecting periodic signals.
We use $P_{\rm orb}$ and $P_{\rm SH}$ for the orbital period
and superhump period, respectively.  We introduce the fractional
superhump excess in the frequency unit
$\varepsilon^* \equiv 1-P_{\rm orb}/P_{\rm SH}$.
$\varepsilon^*_+$ and $\varepsilon^*_-$ represent $\varepsilon^*$
for positive and negative superhumps, respectively.

\begin{figure}
  \begin{center}
%    \FigureFile(80mm,110mm){j1922spec2d.eps}
    \FigureFile(88mm,110mm){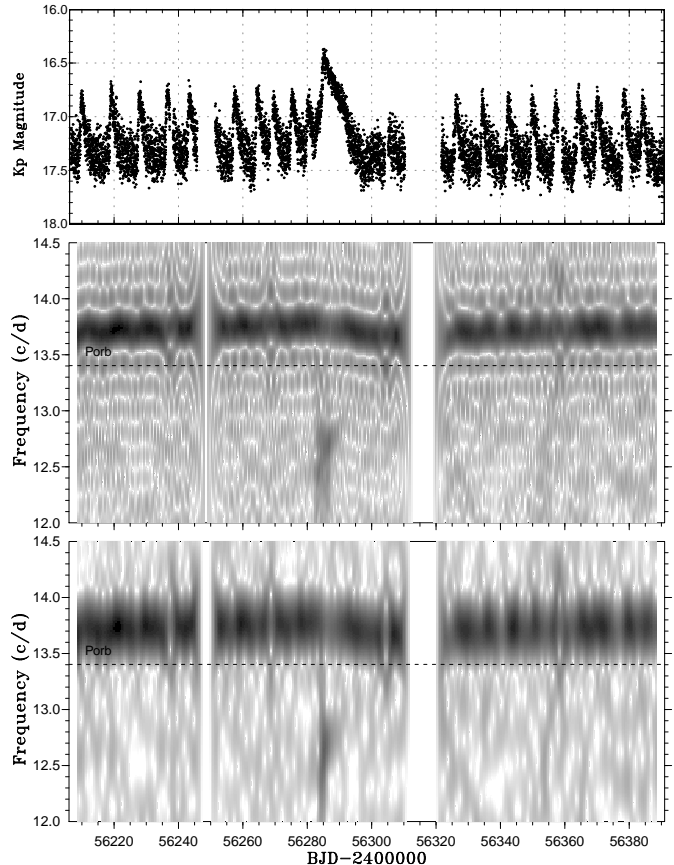}
  \end{center}
  \caption{Two-dimensional power spectrum of the Kepler long cadence
  light curve of KIS J1922.  The orbital period (0.074606~d)
  estimated in subsection \ref{sec:Porb} is marked on the figure.
  (Upper:) Kepler Light curve.
  (Middle:) Power spectrum.  The width of 
  the sliding window and the time step used are 5~d and 0.5~d,
  respectively.  No window function was used.
  (Lower) Power spectrum with Han window function.
  Although the side lobes are suppressed, the frequency resolution
  becomes lower.}
  \label{fig:j1922spec2d}
\end{figure}

   As seen in the upper panel of figure \ref{fig:j1922spec2d},
the Kepler data immediately indicates the pattern of a frequently
outbursting SU UMa-type dwarf nova: the existence of numerous
short outbursts and one long outburst.
The results of the Fourier analysis are shown in 
figure \ref{fig:j1922spec2d}.
Astonishingly, the signal around the frequency 13.7--13.8~c/d
persisted with almost the constant strength (in flux unit) 
throughout the Kepler observation.  Since the frequency is variable,
this signal cannot be the orbital period.  Instead, the frequency
gradually increased prior to the long outburst, and more rapidly
decreased during the long outburst, reached the minimum after
the long outburst and started to increase again gradually.
This pattern of the frequency variation follows the same
pattern of the persistent negative superhumps in V1504 Cyg
(\cite{osa13v1504cygKepler};
\cite{osa13v344lyrv1504cyg}) and V344 Lyr(\cite{osa13v344lyrv1504cyg}).  
Furthermore, we could detect
a transient signal around frequency 12.5--12.8~c/d during
the long outburst (BJD 2456283--2456290).  By analogy with
V1504 Cyg and V344 Lyr, we identified this transient signal
as the positive superhump. Thus the long outburst is indeed 
a superoutburst and this star is an SU UMa star. 

\begin{figure}
  \begin{center}
%    \FigureFile(80mm,110mm){j1922lasso.eps}
    \FigureFile(88mm,110mm){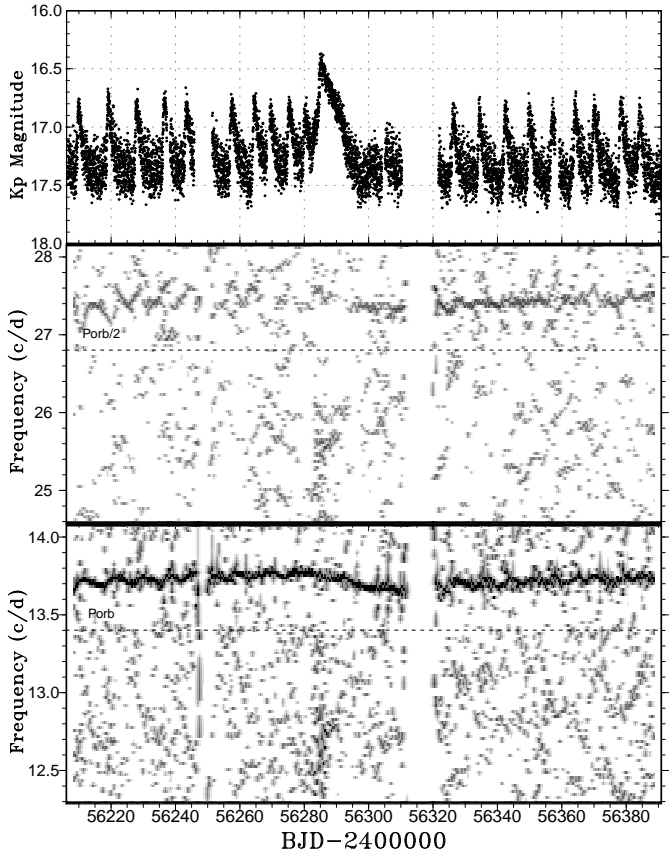}
  \end{center}
  \caption{Two-dimensional Lasso power spectrum of the Kepler
  long cadence light curve of KIS J1922.
  (Upper:) Kepler Light curve.
  (Middle:) Lasso power spectrum ($\log \lambda=-4.5$) 
  of the first harmonic.  The width of 
  the sliding window and the time step used are 4~d and 0.5~d,
  respectively.  The Lasso spectrum was obtained by assuming two
  frequency windows covering the fundamental and the first harmonic.
  The first harmonic component of the negative superhump is stronger
  in the earlier part of the supercycle (i.e., in the later part of the 
  figure, especially after the superoutburst).
  The harmonic component of the positive superhump can be also seen.
  (Lower) Lasso power spectrum of the fundamental.
  The frequency variation of the negative superhump is very clearly
  depicted.  The frequency gradually increased during a sequence of
  normal outbursts.  The frequency also shows a variation in
  accordance with the normal outburst.}
  \label{fig:j1922lasso}
\end{figure}

   The Lasso two-dimensional power spectrum is shown in
figure \ref{fig:j1922lasso}.  This analysis used two
frequency windows covering the fundamental and the first harmonic
and suppressed the signal at other frequencies.
Both the fundamental and the first harmonic signals of
the negative superhumps were clearly detected with
frequency variations as will be discussed below.
The positive superhump was detected as a broad band.
The first harmonic was more strongly detected for the positive
superhump due to strongly non-sinusoidal profile of
the positive superhump.

\section{Discussion}

\subsection{Frequency Variation of Negative Superhumps}\label{sec:NSHvar}

   The global frequency variation of the negative superhumps
closely followed the pattern observed in V1504 Cyg and V344 Lyr.
This global variation reflects the secular increase of 
the disk radius (and the angular momentum) with the progress
of the supercycle.  In addition to this, the frequency reached
a local peak around the maximum of each normal outburst and then
decreased during the subsequent fading.  This pattern is also
common to V1504 Cyg and V344 Lyr, and it reflects the variation
of the disk radius due to the thermal instability.\footnote{
   \citet{sma13negSH} claimed that this interpretation is
   unjustifiable.  We presented our detailed accounts to all of 
   his criticisms by offering clear explanations to his criticisms
   in \citet{osa13v1504cygv344lyrpaper3}, and indicated that
   the variation of the disk radius most strongly affects the
   frequency variation of the negative superhumps.  We followed
   this interpretation throughout this letter.
}
In both respects, we can add yet another example supporting
the TTI model.

   The degree of variation of the frequency in accordance with
the normal outburst is smaller in KIS J1922 than in V1504 Cyg
and V344 Lyr [this becomes more apparent with the phase dispersion
minimization (PDM; \cite{PDM}) analysis in subsection \ref{sec:Porb}].
This can be understood as an effect of a shorter
interval between normal outbursts in KIS J1922: the time
for the disk to shrink is shorter in KIS J1922 (see a discussion
in \cite{osa13v1504cygv344lyrpaper3}. for more potential factors).
KIS J1922 resembles ER UMa and BK Lyn \citep{osa13v1504cygv344lyrpaper3}.
in the small degree of frequency variations in accordance with
the normal outburst interval. 

\subsection{Profile of Negative Superhumps}

\begin{figure}
  \begin{center}
%    \FigureFile(75mm,100mm){j1922profcomp.eps}
    \FigureFile(88mm,110mm){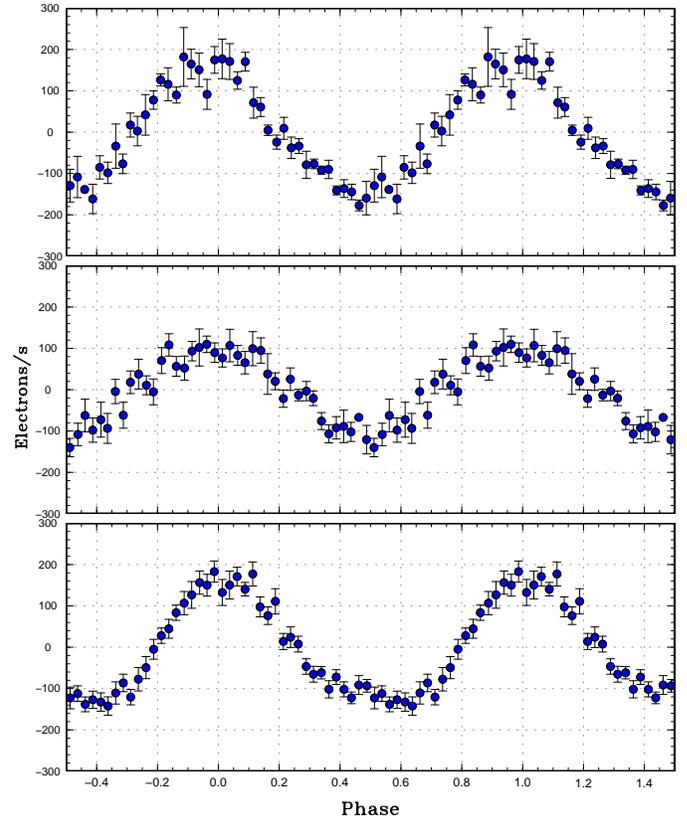}
  \end{center}
  \caption{Profile variation of negative superhumps in KIS J1922.
  (Upper:) Before the superoutburst (BJD 2456250--2456256,
  $P$=0.07272~d).
  (Middle:) During the superoutburst (BJD 2456285--2456292,
  $P$=0.07281~d).
  (Middle:) After the superoutburst (BJD 2456298--2456311,
  $P$=0.07311~d).
  The flux unit 1800 electrons s$^{-1}$ roughly corresponds to 
  the Kepler magnitude of 17.  For a 17.0-mag object,
  the relative amplitude of 200 electrons s$^{-1}$ corresponds to
  an amplitude of 0.11 mag.
  }
  \label{fig:j1922profcomp}
\end{figure}

   The profile of negative superhumps in different phases
is shown in figure \ref{fig:j1922profcomp}.  The profile
tends to have a rather flat minimum and a slightly slower
rise than the fade, all of which are common to the negative
superhumps in V1504 Cyg (figure 6 in \cite{osa13v344lyrv1504cyg}),
although the asymmetry is less prominent in KIS J1922.
This is probably because the disk in KIS J1922 always stays 
close to the hot state, and the profile may not vary strongly
in accordance with the outburst state.

\subsection{Positive Superhumps}

   As seen in figures \ref{fig:j1922spec2d} and \ref{fig:j1922lasso},
the positive superhumps were only briefly
present during the superoutburst, and the strength of the signal
of the negative superhumps always dominated over the positive
superhumps.  In ER UMa, negative superhumps were very prominently
seen even during superoutbursts in 2011--2012 \citep{ohs12eruma}.
Even in the case of ER UMa, the initial stage of the superoutbursts
was dominated by the positive superhumps. 
KIS J1922 is the first object in which the negative superhump 
dominates throughout the supercycle.  Nevertheless, there has been
no exception to the rule that the positive superhumps always appear
during the superoutburst including the most extreme cases such
as the object and ER UMa \citep{ohs12eruma}.
We can conclude that the faithful existence
of the positive superhumps during the superoutburst indicates
that the tidal eccentric instability is essential for triggering
a superoutburst, as predicted by the TTI theory.

\begin{figure}
  \begin{center}
%    \FigureFile(80mm,40mm){j1922so.eps}
    \FigureFile(88mm,50mm){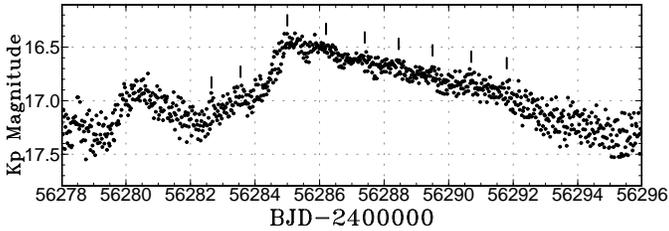}
  \end{center}
  \caption{Enlargement of the superoutburst.  The maxima of
  the beat phenomenon between the positive and negative superhumps
  are labeled by the ticks.}
  \label{fig:j1922so}
\end{figure}

   By looking at the pattern of the beat phenomenon between
the negative and positive superhumps in the light curve
(figure \ref{fig:j1922so}),
we could trace the development of the positive superhumps.
The beat phenomenon with a period of $\sim$1~d could be
detected throughout the plateau part of the superoutburst
with decreasing amplitudes.  This suggests that the positive
superhump were decaying during the plateau phase, and almost
disappeared at the end of the plateau phase.
This behavior is in agreement with the two-dimensional power
spectral analysis, which has a lower time resolution.
The beat phenomenon could be traced to the slowly rising part
(BJD 2456283--2456285) just preceding the superoutburst.
This suggests that the positive superhumps were growing
just after the normal outburst preceding the superoutburst
(i.e. stage A superhumps, \cite{Pdot}).
The normal outburst was thus a precursor outburst as in
V1504 Cyg (BJD 2455876--2455879, \cite{osa13v1504cygv344lyrpaper3}).
The apparent lack of the precursor part in the main
superoutburst (starting from BJD 2456285) can naturally be understood.

\subsection{Estimation of Orbital Period}\label{sec:Porb}

\begin{figure}
  \begin{center}
%    \FigureFile(80mm,90mm){j1922porb.eps}
    \FigureFile(88mm,110mm){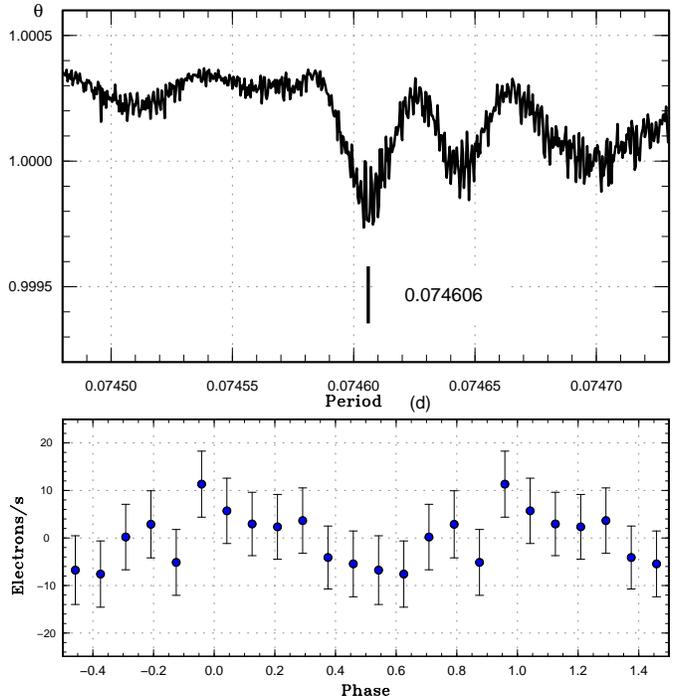}
  \end{center}
  \caption{Possible orbital period.
  (Upper:) PDM analysis.
  (Lower:) Phase-averaged light curve.}
  \label{fig:j1922porb}
\end{figure}

\begin{figure}
  \begin{center}
%    \FigureFile(88mm,70mm){j1922posneg.eps}
    \FigureFile(88mm,70mm){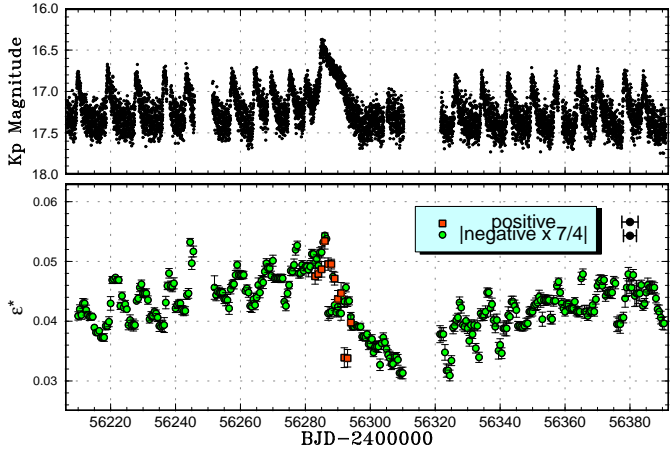}
  \end{center}
  \caption{Variation in precession rates of positive and negative
  superhumps given by two $\varepsilon^*$'s of KIS J1922
  assuming an orbital period of 0.074606~d.
  (Upper:) Light curve.
  (Lower:) Absolute values of fractional superhump excesses
  (positive and negative) in frequency scale determined 
  bt the PDM method.
  The window widths (5~d and 4~d for the positive and negative superhumps,
  respectively; a longer width was adopted for the positive superhumps
  due to the weakness of the signal) are indicated by horizontal bars 
  at the upper right corner and the error bars represent 1$\sigma$
  errors in the frequencies.}
  \label{fig:j1922posneg}
\end{figure}

   As shown in \citet{osa13v344lyrv1504cyg}, $\varepsilon^*_+$
is approximately 7/4 times\footnote{
  This value is weakly dependent
  on the distribution of the mass in the disk, see appendix
  in \citet{osa13v344lyrv1504cyg}.
} larger than $\varepsilon^*_-$
when the pressure effect can be ignored.  By comparing the measured
frequencies of negative and positive superhumps when both signals
are simultaneously present, we can estimate the orbital period.
Since the positive superhumps never became the dominant signal 
in KIS J1922, we consider that the eccentricity is almost 
confined to the outer edge of the disk and the pressure effect
can be ignored (cf. \cite{osa13v344lyrv1504cyg}).
This relation was obtained when we assume an orbital period of
0.07455$\pm$0.00010~d.  We found a candidate orbital period
(5$\sigma$ detection) of 0.074606(1)~d in this period range
using the entire LC data (figure \ref{fig:j1922porb}).
This period well explains the relation between $\varepsilon^*_+$
and $\varepsilon^*_-$ (figure \ref{fig:j1922posneg}) as in
V1504 Cyg and V344 Lyr \citep{osa13v344lyrv1504cyg}.

   This candidate orbital period appears to be sometimes
present in the Lasso spectrum (figure \ref{fig:j1922lasso}).
The weakness of the signal of the orbital period is consistent
with the low orbital inclination suggested from the spectrum.

   Assuming this orbital period, we can estimate the mass ratio
($q=M_2/M_1$) by the method of \citet{kat13qfromstageA}.
Since the positive superhump grew (stage A superhumps)
during BJD 2456281--2456285, we used the frequency of this interval 
[12.77(2)~c/d] and obtained $q$=0.14(1).
The minimum frequency [12.68(2)~c/d] was recorded in the segment
BJD 2456284--2456288.  This frequency corresponds to $q$=0.16(1).
Considering the uncertainties in the frequency determination,
we estimated $q$ to be in a range of 0.14--0.16.

\subsection{Cycle Length and Estimated Supercycle}

   The intervals between the normal outbursts ranged from
5--9~d, the shortest ones were seen just before
the superoutburst.  The cycle length is thus between
the extreme ER UMa-type objects (cycle lengths as short
as 4~d; \cite{kat95eruma}; \cite{rob95eruma};
\cite{pat95v1159ori}) and SU UMa-type dwarf novae with
shortest cycle lengths such as YZ Cnc (8~d).

   Although only one superoutburst was recorded in
the Kepler data, we can estimate the supercycle length by using
the systematic slow increase of the frequency of the negative
superhump as supercycle phase increases.
A supercycle of 140(10)~d
produces a smooth frequency variation versus the supercycle
phase before and after the superoutburst, and we expect
that the supercycle length of KIS J1922 is around this value.
This supercycle length is slightly longer than those of
V1504 Cyg [115(1)~d] and V344 Lyr [114(1)~d].

\subsection{Short Duration of Superoutburst}

   The duration of plateau portion of the superoutburst was 
only 7~d (BJD 2456285--2456292), which is much shorter
than those (10--14~d) of most of SU UMa-type dwarf novae.
The case appears similar to ER UMa \citep{ohs12eruma},
in which the duration of the superoutburst is shorter
when negative superhumps appeared.  This consequence is
naturally understood assuming a disk tilt.  If the disk is
tilted, the gas stream from the secondary reaches the inner
part of the accretion disk and thereby reduces the mass-transfer
rate in the outer part of the disk, which is responsible for
a long-duration superoutburst in high mass-transfer systems
\citep{osa95eruma}.

   In addition to this, the positive superhumps appear to be
suppressed in the presence of the strong negative superhumps.
While in most of SU UMa-type dwarf novae, including ER UMa stars,
the signal of positive superhumps can survive for one or two
cycles of normal outbursts (e.g. \cite{pat95v1159ori};
\cite{sti10v344lyr}; \cite{osa13v344lyrv1504cyg}),
the positive superhump disappeared soon after the plateau
phase in KIS J1922.  It appears that the disk tilt may somehow
suppress the tidal eccentric instability, and this possibility
deserves further investigation.

\subsection{Low Outburst Amplitude}

   Although Kepler has no fixed zero points, the object
was always recorded above 800 count s$^{-1}$, suggesting
that the amplitude of this dwarf nova is unusually low.
The full amplitude in the Kepler data only amounted to 1.0 mag.
The object was listed as an object with a $g$-magnitude of
16.61 in the Kepler Input Catalog and faintest magnitude
recorded in the past survey plates was $B$=18.35 
(USNO YB6 Catalog, unpublished,
referred in NOMAD Catalog).  The majority of the past
records suggests that the object varied within a narrow
range of 16.2--17.3 mag (system close to $V$), confirming
the very low outburst amplitudes in Kepler observations.
There is no contaminating object brighter than magnitude 19
within \timeform{15''} of the object.  An examination of
the Digitized Sky Survey could not resolve any close companion.
Considering that the object is faint in the infrared
($J$=16.94 and $K_s$=17.27 in 2MASS) and the $V-J$ color
is consistent with a CV disk, 
we consider this light comes from the accretion disk,
rather than an unresolved companion.

   The low amplitude suggests that the accretion disk
does not fully return to the true quiescent state (low state)
and the some part of the disk always remains hot.

\subsection{Cycle Length of Normal Outbursts}

   \citet{osa13v1504cygKepler}, \citet{ohs12eruma},
\citet{zem13eruma} indicated that the presence of negative
superhumps suppresses the occurrence of normal outbursts.
This appears to be a reasonable consequence if the disk is
indeed tilted: the mass supply to the outer disk is reduced
and outside-in type outburst is expected to be suppressed.
However, in the case of KIS J1922, the cycle length is
as short as 5--9~d, and all normal outbursts appear to be
outside-in type accompanied by a rapid rise (in $\sim$1~d).
This may be understood if the mass-transfer rate in KIS J1922
is close to the thermal stability, and the disk stays only
slightly below the thermal stability.  In such a system,
the normal outbursts could occur more frequently (with
intervals as short as 4~d) as in ER UMa stars, and the cycle
length of 5--9~d may reflect the reduced occurrence of
normal outbursts in the presence of a disk tilt.
The condition appears to be close to BK Lyn (see also
subsection \ref{sec:NSHvar}).  If the disk in KIS J1922
ceases to tilt, the object may enter the novalike state 
as in BK Lyn (cf. \cite{Pdot5}).

   Only a few systems below the period gap have (nearly)
thermally stable disks.  The well-known examples include BK Lyn
(cf. \cite{pat13bklyn}) and post-novae such as V1974 Cyg
\citep{ski97v1974cyg}.  BK Lyn has recently been proposed to be
a post-outburst nova (\cite{her86novaID}; \cite{pat13bklyn}),
suggesting that all these high mass-transfer objects
below the period gap are post-outburst novae.  Given the unusual
nature of KIS J1922, a search for a nova remnant
(cf. \cite{sha12atcncnova}; \cite{sha12zcamnovashell})
may be fruitful.

\medskip

We thank the Kepler Mission team and the data calibration engineers for
making Kepler data available to the public.
This work was supported by the Grant-in-Aid
``Initiative for High-Dimensional Data-Driven Science through Deepening
of Sparse Modeling'' from the Ministry of Education, Culture, Sports, 
Science and Technology (MEXT) of Japan.

\end{document}